\documentclass[showpacs,preprintnumbers,amsmath,amssymb,twocolumn]{revtex4}

\usepackage{epsfig}

\usepackage{graphicx}

\usepackage{dcolumn}

\usepackage{bm}
\newcommand{\be}{\begin{equation}}

\newcommand{\ee}{\end{equation}}\newcommand{\ba}{\begin{eqnarray}}

\newcommand{\ea}{\end{eqnarray}}

\begin{document}

\title{Diquark-Antidiquarks with Hidden or Open Charm and the Nature of 
$X(3872)$}

\author{L. Maiani}\email{luciano.maiani@roma1.infn.it}
\affiliation{Universit\`{a} di Roma `La Sapienza' and I.N.F.N., Roma, Italy}

\author{F. Piccinini}\email{fulvio.piccinini@pv.infn.it}
\affiliation{I.N.F.N. Sezione di Pavia and Dipartimento di Fisica Nucleare 
e Teorica, Pavia, Italy}

\author{A.D. Polosa}\email{antonio.polosa@cern.ch}
\affiliation{Dipartimento di Fisica, Universit\`{a} di Bari and I.N.F.N., 
Bari, Italy}

\author{V. Riquer}\email{veronica.riquer@cern.ch}
\affiliation{CERN Theory Department, CH-1211, Switzerland}

\date{\today}

\begin{abstract}
Heavy-light diquarks can be the building blocks of a rich spectrum of
states which can accommodate some of the newly observed charmonium-like 
resonances not fitting a pure $c\bar c$ assignment.
We examine this possibility for hidden and open charm diquark-antidiquark
states deducing spectra from constituent quark masses 
and spin-spin interactions.
Taking the $X(3872)$ as input we predict the existence of a $2^{++}$ state
that can be associated to 
the $X(3940)$ observed by Belle and re-examine the state 
claimed by SELEX, $X(2632)$. The possible assignment of 
the previously discovered states $D_s(2317)$ 
and $D_s(2457)$ is discussed.
We predict $X(3872)$ to be made of two components with a mass difference
related to $m_u-m_d$ and discuss the production of $X(3872)$ and of its
charged partner $X^{\pm}$ in the weak decays of $B^{+,0}$.\newline

ROMA1-1396/2004, FNT/T-2004/20, BA-TH/502/04, CERN-PH-TH/2004-239
\pacs{12.40.Yx, 12.39.-x, 14.40.Lb}

\end{abstract}

\maketitle


\section{Introduction}

 It is an old idea that the light scalar mesons $a(980)$ and $f(980)$ may be 
4-quark bound states~\cite{Jaffe}. The idea was more or less 
accepted in the mid-seventies but then it lost momentum, due to 
contradictory results that led the lowest-lying candidate members of a 
diquark-antidiquark nonet, $\sigma$ and $\kappa$, to disappear from the 
Particle Data Tables.

 As an alternative, the possibility was considered that $a(980)$ and $f(980)$ 
may be $K-{\bar K}$ bound states, kept together by hadron exchange forces, 
the same that bind nucleons in the nuclei, color singlet remnants of the 
confining color forces (hence the name $K-{\bar K}$ molecules~\cite{DRGG} used 
in this connection). If they are indeed $K-{\bar K}$ molecules, scalar mesons 
do not need to make a complete SU(3) multiplet so that this idea would be 
consistent with the lack of evidence of light $\sigma$ and $\kappa$. On the 
contrary, since the latter particles would in any case lie considerably higher 
than the respective thresholds, it would be very hard to consider either of 
them as a $\pi-\pi$ or $\pi-K$  molecule. We see that the existence or absence 
of the light scalars is crucial in assessing the nature of $a(980)$ and 
$f(980)$. 

 From this point of view, the recent observations of $\sigma(480)$ and 
$\kappa(800)$ in D non-leptonic decays at Fermilab~\cite{Aitala-K} and in the
 $\pi\pi$ spectrum in $\phi\rightarrow \pi^0\pi^0\gamma$ at 
Frascati~\cite{Aloisio-sig}, have considerably reinforced the case of a full 
nonet with inverted spectrum, as expected for $[qq][{\bar q}{\bar q}]$ states 
and fully antisymmetric diquark ($[qq]: {\rm color}={\bf \bar 3}, 
{\rm flavor}={\bf \bar 3}, {\rm spin}={\bf 0}$). The isolated $I=0$ state is the lightest and it likes to decay in 
$\pi\pi$; the heaviest particles have $I=1, 0$ and like to decay in states 
containing strange quark pairs.

 In the late seventies, diquark-antidiquark mesons have been considered also 
from a different point of view, the so-called 
baryonium~\cite{Rossi&Veneziano77}. Baryonium resonances are 
called for by the extension to baryon-antibaryon scattering of the 
Harari-Rosner duality~\cite{Har}, which is well obeyed by meson-baryon 
amplitudes. In the latter case, the exchange of $q{\bar q}$ mesons in the 
t-channel is dual to (indeed it implies the existence of) non-exotic baryon 
resonances in the s-channel. Similarly, $q{\bar q}$ exchange in the t-channel 
of baryon-antibaryon scattering would give rise to $[qq][{\bar q}{\bar q}]$ 
states in the s-channel.

 In the constituent quark model, emphasis for the binding of the diquark is on 
spin-spin forces, which may lead to strong attraction in the completely 
antisymmetric state. The baryonium picture adds a further element, the 
internal string structure associated with the confining, spin-independent, 
color forces~\cite{Rossi&Veneziano77}. QCD vacuum restricts to a 
one-dimensional string the color lines of force emerging from each 
quark~\cite{'t-Hooft}. In baryons, the strings from the three quarks 
join in a point to form a gauge invariant color singlet and give rise to a 
$Y$-shaped topological structure, with a quark sitting at each ends of the 
$Y$. Decay of baryon resonances is produced by the breaking of one of these 
strings, to give a $q{\bar q}$ meson and a lighter baryon 
(e.g. $\Delta\rightarrow\pi+N)$. In the same picture, duality implies the 
string structure of the diquark-antidiquark states to be that of an $H$, with 
the two quarks sitting on one side and the two antiquarks on the other side of 
the $H$. Topologically, an $H$-shaped state can be seen to arise from the 
fusion of two $Y$-shaped objects, the baryon-antibaryon pair. Conversely, 
OZI allowed decays~\cite{OZI} originating from the breaking of a string 
correspond to decays into either a lighter $qq{\bar q}{\bar q}$ state plus a 
$q{\bar q}$ meson or into baryon-antibaryon. For the lightest scalar mesons 
of each flavor these channels are forbidden by energy conservation and we 
expect basically narrow states. The decay into meson-meson pairs has to 
proceed via the tunneling of the $H$-shaped configuration into two particles
with quarks and antiquarks joined by a single 
string: $S\rightarrow (q{\bar q})+(q{\bar q})$. This picture is 
shown in~\cite{scalar1} to give a reasonable description of the decay 
amplitudes of the lightest scalar mesons.

 If the lightest scalar mesons are diquark-antidiquark composites, it is 
natural to consider analogous states with one or more heavy 
constituents~\cite{scalar1}~\cite{scalar2} (see~\cite{lipkincharm} for an 
early proposal). 

 The aim of the present paper is a study of 
diquark-antidiquark states with hidden or open charm of the form: 
$[cq][{\bar c}{\bar q}^{\prime}]$ and $[cq][{\bar s}{\bar q}^{\prime}]$, 
$q, q'=u, d$. 
 With respect to Ref.~\cite{scalar1}, we add two new elements, the near 
spin-independence of heavy quark forces and isospin breaking from 
quark masses.  We find some unexpected results and predictions, summarized 
in the following.

 For $[cq][{\bar c}{\bar q^{\prime}}]$ states, the approximate 
spin-independence 
of heavy quark interactions~\cite{spinindep}, which is exact in the limit of 
infinite charm mass, implies spin one diquarks to form 
bound states if spin zero diquarks do so 
(``bad'' and ``good'' diquarks, in Jaffe's terminology~\cite{jaffeexotic}).
 A rich spectrum is implied, with states 
with  $J=0, 1, 2$ and both natural and unnatural $J^{PC}$. We describe the 
mass spectrum in terms of (i) the constituent diquark mass and (ii) spin-spin 
interactions. We derive the strength of the latter interactions from the 
known meson and baryon spectrum, where possible, or from educated guesses 
from one-gluon exchange, otherwise. 

 We identify the $X(3872)$~\cite{x3872} with the $J^{PC}=1^{++}$ state with 
the symmetric spin distribution 
$[cq]_{S=1}[{\bar c}{\bar q}]_{S=0}+[cq]_{S=0}[{\bar c}{\bar q}]_{S=1}$ 
(the charmonium assignment and its difficulties are described in
Ref.~\cite{quigg}).
Our assignment is consistent with the observed decays into charmonium plus 
vector mesons. It also implies a pure $S=1$ configuration for the $c-{\bar c}$ 
pair, thus complying with the selection rules derived in Ref.~\cite{voloshin}. 
We have one $J^{PC}=2^{++}$ state at $3952$~MeV that, within the accuracy of 
the model, could be 
identified with the $X(3940)$ seen in Belle data~\cite{belleX2}. 
The scheme features two, $J^{PC}=0^{++}$, states not yet identified. One, 
at $3830$~MeV, 
could decay into $D-{\bar D}$ while the other is below the $D-{\bar D}$ 
threshold and should decay into $\eta_c$+ ps mesons or multihadron 
states. Finally, there are two 
$J^{PC}=1^{+-}$ levels, predicted around $3760$~MeV and $3880$~MeV, also not yet seen.   
 
 It is unclear to us if ``bad'' diquarks with light flavors can bind to 
$[{\bar c}{\bar q}]$, let alone to a completely light-flavored antidiquark, or 
if the stronger repulsion in the $S=1$ state will suppress bound state 
formation. In the first case, an even richer spectrum is implied, 
due to the flavor symmetry of the light diquark, with many exotic states. 

 We extend the previous calculation to the spectrum of 
$[cq][\bar s \bar q^{\prime}]$ states, which can be computed on the basis of 
the same parameters. The resulting spectrum can accommodate the $X(2632)$
claimed by the SELEX Collaboration~\cite{selex}, 
as well as two other states 
previously discovered,  namely 
$D_{s}(2317)$ (with $J^P=0^+$) and $D_{s}(2457)$ ($J^P=1^+$)~\cite{Dsj},
which could be at odds with a $c \bar s$ assignation~\cite{cinesi}.
The latter hypothesis has been discussed in various papers~\cite{ciesse}.
Also a molecular composition of these states has been taken in
consideration, see for example Ref.~\cite{close}.
For a review and a more extended collection of references 
on this topic see e.g.~\cite{stone}.
 
 At the large momentum scales implied by the heavy quark,  the strength of 
self-energy annihilation diagrams decreases. As a consequence, particle masses should 
be approximately diagonal with quark masses, {\it even for the up and down 
quarks}~\cite{scalar2,RossiVen}. Neglecting annihilation diagrams, 
the neutral mass eigenstates 
coincide with: 
\begin{equation}
X_u=[cu][\bar c \bar u];\; X_d=[cd][{\bar c}{\bar d}]
\label{eq:Xud}
\end{equation}
Deviations from this ideal situation is described by a mixing angle 
between $X_u$ and $X_d$. Considering the higher (h) and lower (l) 
eigenvalues, we predict: 
\begin{eqnarray}
&&M(X_h)-M(X_l)=2(m_{d}-m_{u})/\cos(2\theta)=  \nonumber  \\
&&=(7\pm 2)/\cos(2\theta) \; {\rm MeV} 
\label{eq:massdelta}
\end{eqnarray}
in terms of the up and down quark mass difference~\cite{quarkmass}.

 Isospin is broken in the mass eigenstates and, consequently, in their strong 
decays. In particular, we expect this to be the case for $X_h$ and $X_l$, 
which are predicted to decay into both $J/\Psi + \rho$ and 
$J/\Psi + \omega$, as indeed seems to be the case~\cite{belleispin} for 
$X(3872)$. A precise measurement of the branching ratios can provide a determination
 of $\sin \theta$ and therefore a precise prediction of the mass difference.

 We analyze, in this context, the process in which the light vector 
meson from $X$ decay goes into a lepton pair, with $\rho-\omega$ interference, which 
allows to distinguish between the two states $X_h$ and  $X_l$.     

 Finally, we analyze the non-leptonic decay amplitudes $B\rightarrow K X$, for 
both $B^+$ and $B^0$, restricting for simplicity to zero mixing. 

 From the limit to the width of $X(3872)$ as observed by 
Belle~\cite{x3872}, we infer that only one particle should 
dominate the final state 
of $B^+$, either $X_u$ or $X_d$. The $\Delta I=0$ rule of the weak 
transition implies then that $B^0$ decay is dominated by the other state, 
$X_d$ or $X_u$: {\it a precise measurement of the $X$ mass in $B^+$ and $B^0$ 
decay should reveal the mass difference given in (\ref{eq:massdelta})}. 
The observation of the 
decays $X\rightarrow J/\Psi+e^+ e^-$, mentioned in the previous paragraph, 
would 
allow an independent check of which particle is which in $B^+$ and $B^0$ 
decays. We derive also bounds for the production of the charged states 
$X^{\pm}$ in $B$ decays, which are close to, but not in conflict with 
the negative results published 
by 
BaBar~\cite{noXpmbabar}.

It is hardly necessary to remark that our scheme 
is alternative to the 
$D-D^*$ molecule picture proposed for the $X(3872)$~\cite{Dmolecule}. Albeit 
in some case one gets similar predictions (like isospin breaking decays) the 
particle content and the pattern of predictions is quite different, in a way 
that we believe can be put to a test in the near distant future.  

 The plan of the paper is the following. We discuss in Sect. II spin-spin 
interactions in the constituent model and in Sects. III and IV the 
spectrum of the 
hidden and open charm states. Sect. V is devoted to isospin breaking, in 
Sect. VI we discuss the $X$ decays. The 
production of $X$ states in non-leptonic decays of $B^{+,0}$ is discussed in 
Sect. VII. We present our conclusions in Sect. VIII. 

\section{Constituent quarks and spin-spin interactions}  

 In its simplest terms, the constituent quark model~\cite{DRGG,CQothers} 
derives hadron masses from three ingredients: quark composition, constituent 
quark masses and spin-spin interactions. The Hamiltonian is:
\begin{equation}
H=\sum_{i} m_i+\sum_{i<j} 2 \kappa_{ij}(S_i\cdot S_j)
\label{eq:hamilt}
\end{equation}
and the sum runs over the hadron constituents. The coefficients $\kappa_{ij}$ 
depend on the flavor of the constituents $i,j$ and on the particular color 
state of the pair. 

 It is not at all clear how this simple {\it Ansatz} can be derived from the 
basic QCD interaction, in particular how comes that the effect of the 
spin-independent color forces, responsible for quark confinement, can be 
summarized additively in the constituent masses. However, it is a fact 
that Eq.~(\ref{eq:hamilt}) describes well the spectrum of mesons and 
baryons, 
with approximately the same values of the parameters for different situations. 
The spin-spin interaction coefficients scale more or less as expected with 
constituent masses and, when compared in different color states, with the 
values of the color Casimir coefficients derived from one-gluon exchange (as 
we shall see, this is less accurate). Be as it may, we shall accept the 
simple Hamiltonian of Eq.~(\ref{eq:hamilt}). The rest of the Section 
is devoted 
to determining the parameters from the meson and baryon masses. We 
summarize the 
(well known) mass formulae 
for the $s \bar q$ and $s q$ pairs (throughout the paper: $q=u,d$) and 
give a summary of the parameters in Tables I to III.

\begin{table}
\begin{center}
\begin{tabular}{|l|l|l|l|l|}
\hline
  &  $q$ & $s$ & $c$  \\  \hline
constituent & 305    & 490          & 1670 \\\cline{2-4}
 mass (MeV)     & 362      & 546          & 1721 \\
\hline
\hline
\end{tabular}
\end{center}
\caption{Constituent quark masses derived from the $L=0$ mesons (first row)
or from the $L=0$ baryons (second row).}
\end{table}

\begin{table}
\begin{center}
\begin{tabular}{|l|l|l|l|l|l|l|l|l|}
\hline
& $q\bar q$ & $s\bar q$ &$s\bar s$ & $c\bar q$ & $c\bar s$ & $c \bar c$  \\  
\hline
$(\kappa_{ij})_{\bf 0}$~(MeV) & 315 & 195 & $121^*$ & 70 & 72 & 59  \\
\hline
$(\kappa_{ij})_{\bf 0} m_i m_j$(GeV)$^3$ & 0.029 & 0.029 &  & 0.036 & 0.059& 0.16 \\
\hline
\hline
\end{tabular}
\end{center}
\caption{Spin-spin couplings for quark-antiquark pairs in color singlet
from the hyperfine splittings of $L=0$ mesons (first row). The values in the 
second row show the approximate scaling of the couplings with inverse masses 
(masses from meson spectrum). 
*The $s\bar s$ coupling which is not 
experimentally accessible, is obtained by rescaling the $s\bar q$ one by the
factor $m_q/m_s$.}
\end{table}
\begin{table}
\begin{center}
\begin{tabular}{|l|l|l|l|l|l|l|l|l|}
\hline
& $qq$ & $sq$ & $cq$ & $cs$  \\  \hline
$(\kappa_{ij})_{\bf \bar 3}$~(MeV) & 103 & 64& 22& 25 \\
\hline
$(\kappa_{ij})_{\bf \bar 3} m_i m_j$(GeV)$^3$ & 0.014 & 0.013 & 0.014& 
0.024 \\
\hline
\hline
\end{tabular}
\end{center}
\caption{Spin-spin couplings for quark-quark pairs in color 
${\bf \bar 3}$ state
from $L=0$ baryons. One gluon exchange implies 
$(\kappa_{ij})_{\bf \bar 3}=1/2 (\kappa_{ij})_{\bf 0}$. 
The values in the second
row, show the approximate scaling of the couplings with inverse 
masses (masses from the baryon spectrum).}
\end{table}

 Applied to the $L=0$ mesons, $K$ and $K^*$, Eq.~(\ref{eq:hamilt}) 
gives~\cite{pdg}:

\begin{equation}
M=m_q+m_s+\kappa_{s{\bar q}}\left[J(J+1)-\frac{3}{2}\right]
\label{eq:mesons}
\end{equation}
Adding the analogous equations for $\pi-\rho$, $D-D^*$, 
$D_s-D_{s}^*$, we find four 
relations for the constituent masses and the values 
of the four couplings, as reported in Tables I and II. 
There is one consistency 
condition for the constituent masses, which can be written as:
\begin{eqnarray}
&&(m_c+m_q)_D+(m_s+m_q)_K-(2m_q)_{\pi} =  2157\; \; {\rm MeV} \nonumber \\
&& (m_c+m_s)_{D_s} = 2076\; \; {\rm MeV}
\label{eq:massconsist1}
\end{eqnarray}
This relation is representative of the inaccuracy of the model. Spin-spin 
interactions scale as expected, like the inverse product of the masses of 
participating quarks, a most remarkable feature.

 Adding the $J/\Psi-\eta_c$ complex, we obtain the $c{\bar c}$ coupling, also 
reported in Table II, and a considerably smaller constituent charm mass:
\begin{equation}
(m_c)_{J/\Psi}= 1534\; \; {\rm MeV}
\label{eq:massJPsi}
\end{equation}
The parameters from the $J/\Psi$ system deviate appreciably from the rest, 
a not unexpected feature since the charmonium wave function is determined 
by the charmed quark mass and therefore is considerably different from those 
of the mixed 
flavor mesons, which are determined by the light quark masses.

 Baryon masses allow us to obtain quark-quark spin interaction in a color 
antitriplet state. We consider the $uds$ states, $\Lambda$ (``good'' 
diquark, $S=0$), $\Sigma$ and $Y^*$ (``bad'' diquark, $S=1$). One finds:

\begin{eqnarray}
M(S, J)&=&2m_q+m_s+(\kappa_{qq})_{\bf \bar 3}\left[S(S+1)-\frac{3}{2}
\right] \nonumber \\
&+&(\kappa_{qs})_{\bf \bar 3}\left[J(J+1)-S(S+1)-\frac{3}{4}\right]
\label{eq:baryons}
\end{eqnarray}

We can write similar equations for $P-\Delta^+$, involving only 
($\kappa_{qq})_{\bf \bar 3}$, and for $\Lambda_c$, $\Sigma_c$, 
$\Sigma_{c}^*$, involving $(\kappa_{qq})_{\bf \bar 3}$ and 
$(\kappa_{qc})_{\bf \bar 3}$. We find three determinations of 
$(\kappa_{qq})_{\bf \bar 3}$, the values of $(\kappa_{qs})_{\bf \bar 3}$ 
and $(\kappa_{qc})_{\bf \bar 3}$, as well as a new determination 
of the three constituent masses. The new information is given in 
Tables I and III. The consistency conditions read:
\begin{eqnarray}
&&(\kappa_{qq})_{\bf \bar 3}(P-\Delta)=97\; \; {\rm MeV} \nonumber \\
&&(\kappa_{qq})_{\bf \bar 3}(\Lambda-\Sigma-\Sigma^*)=103\; \; {\rm MeV} 
\nonumber \\
&&(\kappa_{qq})_{\bf \bar 3}(\Lambda_c-\Sigma_c-\Sigma_{c}^*)=107\; \; {\rm MeV}
\label{eq:kappaconsist1}
\end{eqnarray} 
For completeness, we consider also the three $\Xi_c$ states, involving 
$(\kappa_{qs})_{\bf \bar 3}$ again and $(\kappa_{sc})_{\bf \bar 3}$. From the 
masses, we find:

\begin{eqnarray}
&&(\kappa_{qs})_{\bf \bar 3}(\Xi_c)=78\; \; {\rm MeV} \nonumber \\
&&(\kappa_{sc})_{\bf \bar 3}(\Xi_c)=25\; \; {\rm MeV} 
\label{eq:kappaconsist2}
\end{eqnarray}

 The overall agreement is quite satisfactory, in particular for the spin-spin 
couplings. The decreasing strength with increasing mass is evident, until we 
go to $cs({\bar s})$ or $c{\bar c}$ states which may have considerable 
distortions in their wave functions.

 For our purposes, however, we need to consider further couplings, which 
refer to the quark-antiquark interactions to which 
we have not yet experimental access. Inside our states, these pairs are 
in a superposition of color singlet and color octet. Omitting spinor 
and space time variables,we write:
\begin{eqnarray}
[cq][{\bar c}{\bar q}]&=&\epsilon^{abc}\epsilon_{ab'c'}(c_b q_c)
({\bar c}^{b'}{\bar q}^{c'}) \nonumber \\
&=&(c_b q_c)({\bar c}^{b}{\bar q}^{c})-(c_b q_c)({\bar c}^{c}{\bar q}^{b})
\label{eq:color1x}
\end{eqnarray} 
Color indices in the last term of Eq.~(\ref{eq:color1x}) can be rearranged 
with the use of the familiar color Fierz-identities:
\begin{equation}
\sum_a \lambda^a_{ij}\lambda^a_{kl}=2\left(\delta_{ik}\delta_{lj}-
\frac{1}{N_c}\delta_{ij}\delta_{lk}\right)\nonumber,
\end{equation}
to put into evidence the 
state of color of the $c{\bar c}$ pair:
\begin{eqnarray}
[cq][\bar c \bar q]=\frac{2}{3}(c_b q_c)({\bar c}^b
{\bar q}^c)-\frac{1}{2}({\bar c}\lambda^A c)({\bar q}\lambda^A q)
\label{eq:color1}
\end{eqnarray}
 
 It is not difficult from Eq.~(\ref{eq:color1}) to see 
that the probability to find a  particular $q \bar q$ pair in color octet 
is twice the 
probability of the color singlet, so that (the same holds for the other 
flavors as well):

\begin{equation}
\kappa_{c\bar c}([cq][\bar c \bar q])=\frac{1}{3}
(\kappa_{c \bar c})_{{\bf 0}} 
+ \frac{2}{3}(\kappa_{c\bar c})_{{\bf 8}}
\label{eq:color2}
\end{equation}

Of course, we do not know $(\kappa_{c \bar c})_{\bf 8}$. We resort 
to the rule derived from one-gluon exchange:

\begin{equation}
(\kappa_{c\bar c})_{\bf X}={\rm const.}[C^{(2)}
({\bf X})-C^{(2)}({\bf 3})-C^{(2)}
({\bf \bar 3})]
\label{eq:color3}
\end{equation}
where $C^{(2)}(X)$ is the value of the quadratic 
Casimir operator in the representation ${\bf X}$: 
$C^{(2)}({\bf X})= 0, 3, 4/3, 4/3$ for ${\bf X}={\bf 0}, {\bf 8}, {\bf 3}, 
{\bf \bar 3}$. 
Eqs.~(\ref{eq:color2}) and (\ref{eq:color3}) give, in conclusion:

\begin{equation}
\kappa_{c\bar c}=\kappa_{c\bar c}([cq][{\bar c}{\bar q}])=\frac{1}{4}
(\kappa_{c\bar c})_{\bf 0}
\label{eq:color4}
\end{equation} 
 
 We apply the previous results to determine the constituent mass of light 
diquarks, considering explicitly the case of the $a_0(980)$:
\begin{equation}
a_0(980)=[sq]_{S=0}[{\bar s}{\bar q}]_{S=0}
\label{eq:azero}
\end{equation}

We write the Hamiltonian according to:

\begin{eqnarray}
H=2m_{[cs]}&+&2(\kappa_{sq})_{\bf \bar 3}[(S_s \cdot S_q)+(S_{\bar s} 
\cdot S_{\bar q^{\prime}})] \nonumber \\
&+& 2\kappa_{q{\bar q}}(S_q \cdot S_{\bar q^{\prime}}) \nonumber \\
&+& 2\kappa_{s{\bar q}}[(S_s \cdot S_{\bar q^{\prime}})
+(S_{\bar s} \cdot S_q)] 
\nonumber \\
&+& 2\kappa_{s{\bar s}}(S_s \cdot S_{\bar s})
\label{eq:hamilts}
\end{eqnarray}
 
 The state given in Eq.~(\ref{eq:azero}) is not an eigenstate of this 
Hamiltonian, which is diagonalized only within the states with different 
diquark spin composition, see Sect III below. However, the latter could as 
well not exist, so we content ourselves with the mean value:
\begin{equation}
\langle a_0|H|a_0\rangle = 984\; \; {\rm MeV} = 2m_{[sq]} 
- 3(\kappa_{sq})_{\bf \bar 3}
\label{eq:hamiltlight}
\end{equation} 
and, using the value in Table III, we find:

\begin{equation}
m_{[sq]}=590\; \; {\rm MeV}
\label{eq:msq}
\end{equation}

For $\sigma(480)$, with $(\kappa_{qq})_{\bf \bar 3}$, we find:

\begin{equation}
m_{[ud]}=395~{\rm MeV}
\label{eq:mass[ud]}
\end{equation}
 Light diquarks constituent are not at all much heavier than constituent 
quarks.

\section{The spectrum of $[cq][\bar c \bar q^{\prime}]$ states}

 States can be conveniently classified in terms of the diquark and antidiquark 
spin, $S_{cq}$, $S_{{\bar c}{\bar q}'}$, total angular momentum, $J$, parity, 
P, and charge conjugation, C. We have the following states.

{\bf i.} Two states with $J^{PC}=0^{++}$:
\begin{eqnarray}
&&|0^{++}\rangle = |0_{cq}, 0_{\bar c \bar q^{\prime}};J=0\rangle; 
\nonumber \\ 
&&|0^{++\prime}\rangle= |1_{cq}, 1_{\bar c \bar q^{\prime}};J=0\rangle
\label{eq:zeropp}
\end{eqnarray}

{\bf ii.} Three states with $J=1$ and positive parity:
\begin{eqnarray}
&&|A\rangle=|0_{cq}, 1_{\bar c \bar q^{\prime}};J=1\rangle; \nonumber \\
&&|B\rangle=|1_{cq}, 0_{\bar c\bar q^{\prime}};J=1\rangle; \nonumber \\
&&|C\rangle=|1_{cq}, 1_{\bar c\bar q^{\prime}};J=1\rangle 
\label{eq:onep}
\end{eqnarray}

Under charge conjugation, $|A\rangle$ and $|B\rangle$ interchange while 
$|C\rangle$ is odd. Thus the $1^+$ complex contains one C-even 
and two C-odd states:
\begin{eqnarray}
&&|1^{++}\rangle=\frac{1}{\sqrt{2}}(|A\rangle+|B\rangle);\nonumber\\ 
&&|1^{+-}\rangle=\frac{1}{\sqrt{2}}(|A\rangle-|B\rangle);\nonumber\\
&&|1^{+-\prime}\rangle=|C\rangle
\label{eq:onepp}
\end{eqnarray}
One can analyze these states in terms of the states with definite values 
for the spin of $c{\bar c}$ and $q{\bar q^{\prime}}$. The state with both spins 
equal to zero cannot appear, because $J=1$; among the others, the only one 
with $C=+$ is that with both spins equal to one. Thus, the state 
$|1^{++}\rangle$ in Eq.~(\ref{eq:onepp}) has a definite value of the 
$c{\bar c}$ spin, $S_{c{\bar c}}=1$.

{\bf iii.} One state with $J^{PC}=2^{++}$:
\begin{equation}
|2^{++}\rangle = |1_{cq}, 1_{\bar c \bar q^{\prime}}; J=2\rangle 
\label{eq:twopp}
\end{equation}
The $2^{++}$ state has also $S_{c\bar c}=1$.

 Next, we consider the Hamiltonian, which is the same 
as in Eq.~(\ref{eq:hamilts}) with $s \rightarrow c$:
\begin{eqnarray}
H=2m_{[cq]}&+&2(\kappa_{cq})_{\bf \bar 3}[(S_c \cdot S_q)+(S_{\bar c} 
\cdot S_{\bar q^{\prime}})] \nonumber \\
&+& 2\kappa_{q{\bar q}}(S_q \cdot S_{\bar q^{\prime}}) \nonumber \\
&+& 2\kappa_{c{\bar q}}[(S_c \cdot S_{\bar q^{\prime}})+(S_{\bar c} \cdot S_q)] 
\nonumber \\
&+& 2\kappa_{c{\bar c}}(S_c \cdot S_{\bar c})
\label{eq:hamiltc}
\end{eqnarray}

The Hamiltonian is diagonal on the $1^{++}$ and $2^{++}$ states, with 
eigenvalues:
\begin{equation}
M(1^{++}) = 2m_{[cq]}-(\kappa_{cq})_{\bf \bar 3}+\frac{1}{2}\kappa_{q{\bar q}} 
- \kappa_{c{\bar q}}+\frac{1}{2}\kappa_{c{\bar c}}
\label{eq:single1}
\end{equation}
\begin{equation}
M(2^{++}) = 2m_{[cq]}+(\kappa_{cq})_{\bf \bar 3}+\frac{1}{2}\kappa_{q{\bar q}} 
+ \kappa_{c{\bar q}}+\frac{1}{2}\kappa_{c{\bar c}}
\label{eq:single2}
\end{equation}

 A tedious but straightforward calculation (see Appendix) leads to two, 
$2\times 2$, matrices for the other states.

\begin{eqnarray}
&&M(0^{++})= \nonumber  \\
&&\left( \begin{array}{cc}
-3(\kappa_{cq})_{\bf \bar 3} & \frac{\sqrt{3}}{2}(\kappa_{q\bar q}
+\kappa_{c\bar c}-
2\kappa_{c\bar q}) \\
\frac{\sqrt{3}}{2}(\kappa_{q\bar q}+\kappa_{c\bar c}-2\kappa_{c\bar q})& 
(\kappa_{cq})_{\bf \bar 3}
-(\kappa_{c\bar c}+\kappa_{q\bar q}-2\kappa_{c\bar q})
\end{array}\right)\nonumber
\end{eqnarray}

\begin{eqnarray}
&&M(1^{+-})=\nonumber\\
&&\left( \begin{array}{cc}
-(\kappa_{cq})_{\bf \bar 3} 
+\kappa_{c\bar q}-\frac{(\kappa_{c\bar c}+\kappa_{q\bar q})}{2}
& \kappa_{q\bar q}-\kappa_{c\bar c} \\
\kappa_{q\bar q}-\kappa_{c\bar c} & 
(\kappa_{cq})_{\bf \bar 3}-\kappa_{c\bar q}-\frac{(\kappa_{q\bar q}
+\kappa_{c\bar c})}{2}
\end{array}\right)\nonumber
\end{eqnarray}

The state $1^{++}$ is an almost perfect candidate to explain the properties 
of $X(3872)$:

- it is expected to be narrow, like all diquark-antidiquark systems below the 
baryon-antibaryon threshold;

- the unnatural spin-parity forbids the decay in $D-{\bar D}$, which is not 
observed;

- it can decay in the observed channels 
$J/\Psi+{\rm light \; vector\; meson}$, with 
conservation of the spin of the heavy flavor system;

- it decays into both $\rho$ 
and $\omega$, due to isospin breaking in its wave function (Sect. IV).

 How narrow is narrow we shall consider in Sect. VI. For the moment, we 
identify the $1^{++}$ with the $X(3872)$ and proceed to compute the spectrum 
via the 
couplings of Table III, directly, and those of Table II, scaled according to 
Eqs.~(\ref{eq:color1}), (\ref{eq:color2}) and (\ref{eq:color3}). 
By Eq.~(\ref{eq:single1}), the diquark constituent mass is fixed to be:

\begin{equation}
m_{[cq]}= 1933\;\; {\rm MeV}
\label{eq:mqc}
\end{equation}

 We report in Fig. 1 the full spectrum computed numerically. The energy 
levels have an error which is difficult to quantify at the moment, 
maybe in the order of $10-20$~MeV.
A few observations are in order.

\begin{figure}[ht]
\begin{center}
\epsfig{
height=7truecm, width=8truecm,
        figure=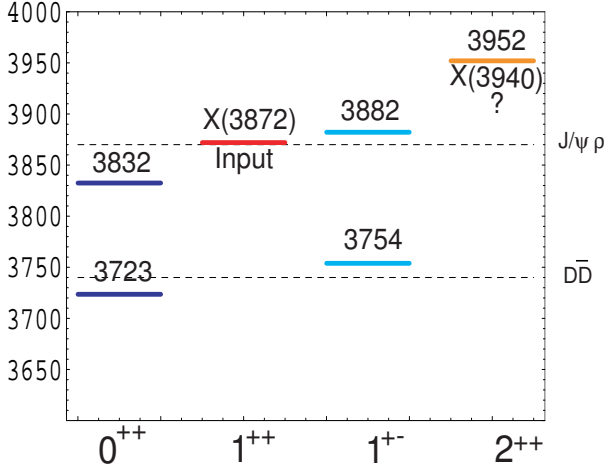}
\caption{\label{fig.1} \footnotesize 
{\small The full spectrum of the $X$ particles.}
}
\end{center}
\end{figure}

\noindent {\bf i.} From 
Eqs.~(\ref{eq:single1}) and (\ref{eq:single2})  we read:

\begin{equation}
M(2^{++}) = M(1^{++}) + 2[(\kappa_{cq})_{\bf \bar 3}+\kappa_{c{\bar q}}]=
3952~{\rm MeV}
\label{eq:mdue}
\end{equation}

This places the $2^{++}$ close to the recently observed~\cite{belleX2} 
resonance at $3940$~MeV. Note that the coupling $\kappa_{cq}$ is well 
determined 
and the other, $\kappa_{c{\bar q}}$, could easily be smaller than we estimate 
with 
the color factor. The identification of the $2^{++}$ with the $X(3940)$ is 
quite attractive. The $2^{++}$ can decay in 
$J/\Psi+ {\rm light\; vector\; meson}$ 
respecting the conservation of the heavy flavor spin and also in $D-{\bar D}$. 
The decay $X(3940)\rightarrow J/\Psi+ \omega$ is seen by Belle, the 
$D-{\bar D}$ decay should be searched for, but it could be somewhat suppressed 
by the decay in D-wave.

\noindent {\bf ii.} Of the two $0^{++}$ states, one is below the $D-{\bar D}$ 
threshold. It can decay in $\eta_c \pi$ or $\eta_c \eta$ or 
multihadron states. The other should be 
seen to decay in $D-{\bar D}$. There are no candidates, at present, for the 
$0^{++}$ states. The same holds for the two $1^{+-}$ states. Allowed decays of 
the latter are $J/\Psi+ \pi(\eta)$, $\eta_c+ \rho(\omega)$.

\section{The  $[cq][\bar s \bar q^{\prime}]$ states}

 We extend the calculation of the previous Section to the states 
$[cq][{\bar s}{\bar q^{\prime}}]$, 
leaving aside the issue whether they can bind or not.
The appropriate Hamiltonian is:
\begin{eqnarray}
&&H=m_{[cq]}+m_{[sq]}+2(\kappa_{cq})_{\bf \bar 3}(S_c \cdot S_q)+\nonumber \\
&&+2(\kappa_{sq})_{\bf \bar 3}
(S_{\bar s} \cdot S_{\bar q^{\prime}}) +\nonumber \\ 
&&+2\kappa_{q \bar q}(S_q \cdot S_{\bar q^{\prime}}) +
2\kappa_{c \bar q}(S_c \cdot S_{\bar q^{\prime}})+ 
2\kappa_{s \bar q}(S_{\bar s} \cdot S_q)+ \nonumber \\
&&+2\kappa_{c{\bar s}}(S_c \cdot S_{\bar s})
\label{eq:hamiltcs}
\end{eqnarray}

The angular momentum composition of the multiplet is, of course, 
the same as the previous one except that this set of states is not invariant 
under $C$-conjugation and the $J^P=1^+$ states form an irreducible complex. 
The energy levels are given by the following formulae.

\begin{eqnarray}
&&M(2^{+}) = m_{[cq]} + m_{[sq]} + \frac{1}{2}[(\kappa_{cq})_{\bf \bar 3}+
+(\kappa_{sq})_{\bf\bar 3}]+ \nonumber \\
&&\;\;\;\;\;\;\;\;\;\;\;\;\;\;+ \frac{1}{2}
\kappa_{q{\bar q}} +\frac{1}{2}(\kappa_{c{\bar q}}+
\kappa_{s{\bar q}})+\frac{1}{2}\kappa_{c \bar s}
\label{eq:2pcs}
\end{eqnarray}

\begin{eqnarray}
&&M(1^+)_{11}=[-3(\kappa_{cq})_{\bf \bar 3}+(\kappa_{sq})_{\bf \bar 3}]/2 \nonumber \\
&&M(1^+)_{12}=(\kappa_{q\bar q}-\kappa_{c \bar q}-\kappa_{s \bar q}+
\kappa_{c\bar s})/2 \nonumber \\
&&M(1^+)_{13}=(\kappa_{q\bar q}-\kappa_{c \bar q}+\kappa_{q \bar s}
-\kappa_{c\bar s})/\sqrt{2} \\
&&M(1^+)_{22}=[(\kappa_{cq})_{\bf\bar 3}-3 (\kappa_{sq})_{\bf \bar 3}]/2 \nonumber \\
&&M(1^+)_{23}=(-\kappa_{q\bar q}-\kappa_{c \bar q}+\kappa_{q \bar s}
+\kappa_{c \bar s})/ \sqrt{2} \nonumber \\
&&M(1^+)_{33}=[(\kappa_{cq})_{\bf\bar 3}+(\kappa_{sq})_{\bf\bar 3}-
\kappa_{q\bar q}-\kappa_{q\bar s}-\kappa_{c\bar q}-\kappa_{c\bar s}]/2 
\nonumber
\end{eqnarray}

\begin{eqnarray}
&&M(0^+)_{11}= -\frac{3}{2}[(\kappa_{cq})_{\bf \bar 3}
+(\kappa_{sq})_{\bf \bar 3}]
\nonumber \\
&&M(0^+)_{12}=\frac{\sqrt{3}}{2}(\kappa_{q\bar q}-\kappa_{c \bar q}-
\kappa_{s \bar q}+\kappa_{c \bar s})  \\
&&M(0^+)_{22}=\frac{1}{2}[(\kappa_{cq})_{\bf \bar 3}+
(\kappa_{sq})_{\bf\bar 3}]-   \nonumber \\
&&\;\;\;\;\;\;\;\;\;\;\;\;\;\;\;\;\;\;-(\kappa_{q\bar q}+\kappa_{c \bar q}+\kappa_{s \bar q}+\kappa_{c \bar s})
\nonumber
\end{eqnarray}

 The spectrum is reported in Fig.~2. 
\begin{figure}[ht]
\begin{center}
\epsfig{
height=6truecm, width=8truecm,
        figure=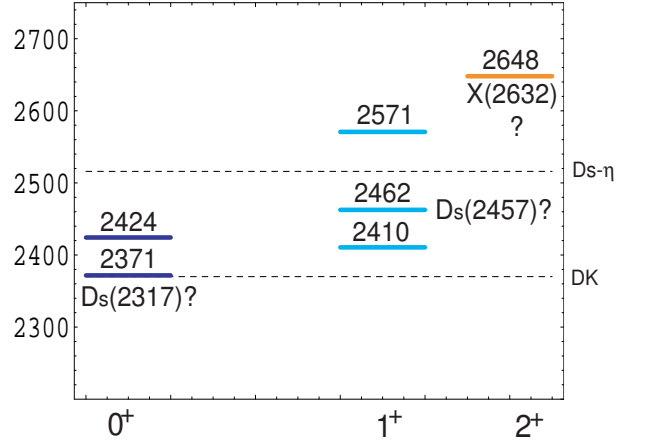}
\caption{\label{fig.2} \footnotesize 
{\small The predicted spectrum of the particles with open charm.}}
\end{center}
\end{figure}

The particle claimed by the SELEX 
Collaboration~\cite{selex} fits quite naturally in it as the $2^+$ member 
of the 
multiplet. Note the change of attribution, with respect to the $0^+$ 
assignment suggested 
previously~\cite{scalar2}, 
which makes the $X(2632)$ compatible with the diquark 
constituent masses found in Eqs.~(\ref{eq:msq})
and~(\ref{eq:mqc}), 
without changing the results presented there. 
 In addition, we associate tentatively the lowest $0^+$  and one 
of the lowest lying $1^+$ with $D_{s}(2317)$ and 
$D_{s}(2457)$~\cite{Dsj}, respectively. 
This is compatible with the observed decays:
\begin{eqnarray}
&&D_{s}(2317) \rightarrow D_s \pi^0; \nonumber \\
&&D_{s}(2457) \rightarrow D_s \gamma \pi^0; \; (D_s)^* \pi^0
\end{eqnarray}

  A four quark interpretation of the $D_{s}$ particles has been advanced 
in Ref.~\cite{cinesi}, while the $c \bar s$ interpretation is pursued 
in~\cite{ciesse}.
 We find quite suggestive that by assigning the $X(3872)$ to its natural, 
$1^{++}$, 
level and using reasonable values of the spin-spin couplings we are 
able to fit other 
four particles, which could be at odds with 
the conventional quark-antiquark 
interpretation.

\section{Isospin breaking}

 We consider in this Section the finer structure of the $X(3872)$. In 
particular, we consider 
the neutral states with the composition given in Eq.~(\ref{eq:Xud}).
Physical states could be expected to fall in isospin multiplets with 
$I=1,0$: 
\begin{eqnarray}
&&f_{c \bar c}=(X_u+X_d)/\sqrt{2}; \nonumber \\
&&a_{c \bar c}=(X_u-X_d)/\sqrt{2}
\label{eq:ispinbasis}
\end{eqnarray}

 The two states in Eq.~(\ref{eq:Xud}) are mixed by self-energy 
diagrams whereby a light quark pair transforms 
into another one by annihilation into intermediate gluons.
In the basis Eq.~(\ref{eq:Xud}) annihilation diagrams contribute 
equally to all entries of the mass matrix. 
The contribution of quark masses, on the other hand, 
is diagonal in the basis Eq.~(\ref{eq:Xud}). The resulting 
$2\times 2$ matrix is:
\begin{eqnarray}
&&\left( \begin{array}{cc}
2m_u+\delta & \delta \\
\delta & 2m_d+\delta
\end{array}\right)\nonumber
\end{eqnarray}
$\delta$ being the contribution from annihilation graphs.
The matrix with all equal  
entries $\delta$, admits the states in (\ref{eq:ispinbasis}) as eigenvectors, 
with split masses. 

 At the mass scale determined by the ${c \bar c}$ pair 
we expect~\cite{scalar2} 
annihilation diagrams to be small, as indicated by the
very small $J/\Psi$ width. Thus, mass eigenvectors should align to 
the {\rm quark mass} basis. For the strange quark, this happens 
already at the mass scale of the vector mesons, $\phi$ and $\omega$. 
 The other case is provided by the  scalars $a(980)$ and $f(980)$,  
which are quite degenerate in mass. The upper bound~\cite{pdg} 
$|\Delta M|<10$~MeV 
indicates that annihilation contributions are, at best, at the 
level of the normal isospin breaking mass differences, suggesting a sizeable 
deviation from the isospin basis.  At the $X(3872)$ scale, we expect 
the $u-d$ quark mass difference to dominate and the mass eigenstates to 
coincide with the states in 
(\ref{eq:Xud}) to a rather good extent. 

A numerical estimate of the mass difference is obtained as follows.
The up and down quark mass difference is determined by the pseudoscalar
meson spectrum~\cite{quarkmass}, after separating its contribution from the 
background of second order electromagnetic corrections due to 
one-photon exchange. 
The so-called Dashen's theorem~\cite{dashen} states 
that one-photon exchange does not contribute to the isospin breaking, U-spin 
singlet combination of Kaon and pion mass differences, which is therefore 
given by the quark mass difference:
\begin{eqnarray}
&&(K^+-K^0)-(\pi^+-\pi^0)=C(m_{u}-m_{d})=\nonumber  \\
&&=-5.3\cdot 10^{-3}\; ({\rm GeV})^2
\label{dash}
\end{eqnarray}
where particle symbols stand for squared masses. Combining with the equations 
for the pion and Kaon masses in terms of quark masses, and assuming 
$m_s=150$~MeV 
from the baryon mass differences, one finds:
\begin{equation} 
(m_{u}-m_{d})= 3.3~{\rm MeV}
\label{eq:dmup}
\end{equation}

 Before translating Eq.~(\ref{eq:dmup}) in hadron mass- differences, one must 
control the one-photon exchange contributions. We can divide the 
e.m. corrections in (i) corrections on the same 
constituent quark line, (ii) photon crossing from one to another quark line. 
In the constituent quark model, the 
first correction goes into a renormalization of the constituent mass while 
the second one adds to the spin-spin interaction. A control case is that of 
the charmed mesons:
\begin{eqnarray}
&&M(D^+)-M(D^0)= 4.78\pm 0.1\;{\rm MeV} \nonumber \\
&&M(D^{*+})-M(D^{*0})= 3.3\pm 0.7\;{\rm MeV}
\label{ddiff}
\end{eqnarray}
Note that the result in the first line is {\it larger} than the mass 
difference in 
(\ref{eq:dmup}): the spin-spin interaction in total spin zero is repulsive 
(attractive ) for non-vanishing (vanishing) total charge. The correction to 
the 
constituent masses is obtained by averaging over the two spin multiplets, see 
Eq.~(\ref{eq:mesons}): 
\begin{eqnarray}
&&(m_{d}-m_{u})_{const}=\frac{3(D^{*+}-D^{*0})+(D^+-D^0)}{4}= \nonumber \\
&&=3.7\pm 0.7 \; {\rm MeV}
\end{eqnarray}
 The agreement with Eq.~(\ref{eq:dmup}) is better than for individual mass 
differences. Unfortunately, at the moment, we do not have enough masses 
to determine and subtract the spin-spin e.m. interaction for the 
$[cq][\bar c \bar q^{\prime}]$ multiplet. We take the example as suggestive
that neglecting photon exchange may introduce an error of, perhaps, 
$\simeq 30\%$. 

 Non-negligible gluon annihilation diagrams mix $X_u$ and $X_d$ and increase 
the mass difference. Writing:
\begin{eqnarray}
&&X_{\rm low}=\cos \theta X_u+\sin \theta X_d\nonumber\\
&&X_{\rm high}=-\sin \theta X_u+\cos \theta X_d
\label{eq:mix}
\end{eqnarray}
we get the result already stated in the Introduction:
\begin{eqnarray}
&&M(X_h)-M(X_l)=2(m_{d}-m_{u})/\cos(2\theta)=  \nonumber  \\
&&=(7\pm 2)/\cos(2\theta) \; {\rm MeV}  \nonumber
\end{eqnarray}

 The mixing angle can be determined from $\Delta M$ as well as from the 
ratio of the 
decay rates in $J/\Psi+\omega$ and $J/\Psi+\rho$, as we shall see in the next 
Section. It goes without saying that the same considerations can be applied
to all states in Figs.~1,2.

 In conclusion, we predict close to maximal isospin breaking in the wave 
function and correspondingly in the hadronic decays of $X(3872)$.

 Isospin violation in the wave function is also predicted by the $DD^*$ 
molecule scheme~\cite{Dmolecule}, with the $X(3872)$ being essentially 
$(D^0 {\bar D^{*0}})+({\bar D^0} D^{*0})$. However, in our scheme we have 
{\rm two states} rather than one, separated by the mass difference 
(\ref{eq:massdelta}), and quite a richer phenomenology.

\section{The $X(3872)$ decay width}

 A pair of color-singlet mesons cannot 
be obtained by cutting the strings that join quarks and antiquarks in the $H$ 
shaped $[qq][\bar q \bar q]$ states. The baryonium picture 
suggests that the two-meson decays of the latter 
go via intermediate 
baryon-antibaryon states of high mass. This implies basically narrow widths. 

 The $X(3872)$ is expected to be particularly 
narrow for several additional reasons. 

\noindent (i) Unnatural spin-parity forbids decays into 
$D \bar D$;
 
\noindent (ii) the channel $D D^*$ is below threshold;
 
\noindent (iii) decay in  $\eta_c+{\rm mesons}$ is forbidden by heavy flavor 
spin conservations~\cite{voloshin}. 

 Of the charmonium channels, the only available ones are $J/\Psi+2\pi$ 
and $J/\Psi+3\pi$, dominated  by $\rho^0$ and $\omega$, respectively. 
Each mass eigenstate decays simultaneously in the two channels, due 
to isospin breaking in the wave function.

 We describe the decay by a single {\rm switch amplitude}, associated to 
the process:
\begin{eqnarray}
&&[cu]_{\bf\bar 3}[\bar c \bar u]_{\bf\bar 3}\rightarrow 
(c \bar c)_{{\bf 0}}(u \bar u)_{{\bf 0}}
\label{eq:switch}
\end{eqnarray} 
where subscripts indicate color configurations. 

 We further write the 
invariant three-meson coupling for $X_u$ according to:
\begin{eqnarray}
{\it L}_{X_u\Psi V}&=&g_V\epsilon^{\mu\nu\rho\sigma}
P_{\mu}X_{\nu}\psi_{\rho}V_{\sigma}=  \nonumber \\
&=& g_V M_{X}({\bf X}\wedge {\bf{\psi}})\cdot{\bf V}
\label{eq:coupling}
\end{eqnarray} 
where $P_{\mu}$ and $M_X$ are the decaying particle momentum and mass. To 
estimate the value of $g_V$, we compare with the similar couplings for
the light scalar mesons, determined by one dimensionful constant, 
$A\simeq 2.6$~GeV. An admittedly bold guess, to obtain the order of 
magnitude, is:
\begin{equation}
g_V M_{X}=\frac{A}{\sqrt{2}}
\label{eq:coupling}
\end{equation} 
 
 By dominating the $2\pi(3\pi)$ decay with $\rho(\omega)$ exchange in 
the narrow-width approximation, we find:
\begin{eqnarray}
&&\frac{d\Gamma(X_l\rightarrow \psi+f)}{ds}= 
\frac{2x_{l,V}|A|^2B_{(V\rightarrow f)}}{8\pi M_X^2}\cdot \nonumber \\
&&\cdot \frac{M_V\Gamma_V}{\pi}
\frac{p(s)}{(s-M_V^2)^2+(M_V\Gamma_V)^2}
\label{eq:breitw}
\end{eqnarray} 
with $f=\pi^+\pi^-(\pi^+\pi^-\pi^0)$ for $V=\rho(\omega)$, $s$ the 
invariant mass-squared of the pions, and $p$ the decay momentum:
\begin{eqnarray}
&&p(s)=\frac{\sqrt{\lambda(M_X,M_{\psi},M_V)}}{2M_X}; \nonumber \\
&&\lambda=(M_X)^4+(M_{\psi})^4+(M_V)^4-2(M_X M_{\psi})^2 \nonumber\\
&&-2(M_X M_V)^2
-2(M_{\psi}M_V)^2
\label{eq:decayp}
\end{eqnarray} 
The coefficient $x_{l,V}$ is:
\begin{eqnarray}
&&x_{l,V}=\frac{(\cos \theta \pm \sin \theta)^2}{2}
\label{eq:coeff}
\end{eqnarray}
for $V=\omega(\rho)$. Similar equations hold for the higher mass state, 
$X_h$, with the appropriate substitutions. 
 
By numerical integration, we then find:
\begin{eqnarray}
&&\langle p\rangle_{\rho}=\left(\frac{M_{\rho}\Gamma_{\rho}}{\pi}\right)
\int_{\rm (2m_{\pi})^2}^{\infty}ds
\frac{p(s)}{(s-M_{\rho}^2)^2+(M_{\rho}\Gamma_{\rho})^2}=\nonumber\\
&&\;\;\;\;\;\;\;\,=126 \; {\rm MeV}; \nonumber\\
&&\langle p \rangle_{\omega}=22 \; {\rm MeV}
\label{eq:pave}
\end{eqnarray}
and:
\begin{eqnarray}
&&\Gamma(X_{l}\rightarrow J/\psi+\pi^+\pi^-)=
\frac{2x_{l,\rho}|A|^2}{8\pi M_X^2}
\langle p\rangle_{\rho}=\nonumber \\
&&=2 x_{l,\rho}\cdot 2.3 \; {\rm MeV}; \nonumber\\
&&\Gamma(X_{l}\rightarrow J/\psi+\pi^+\pi^-\pi^0)=
\frac{2x_{l,\omega}|A|^2}{8\pi M_X^2}
\langle p\rangle_{\omega}=\nonumber\\
&&=2 x_{l,\omega}\cdot 0.4 \; {\rm MeV}
\label{eq:rates}
\end{eqnarray}

 We anticipate small widths, comparable to the resolution of Belle and 
BaBar. However, given the mass difference in (\ref{eq:massdelta}), one would 
expect to observe either 
two peaks or one unresolved structure, broader than the stated experimental 
resolution, $4.5$~MeV. Taking Belle data at face 
value, we conclude that only one of the two neutral states is produced 
appreciably in $B^+$ decay (this will be discussed in the next Section). 
Assuming this to be the case, we can get some information on the mixing 
angle from the observed ratio of $3\pi$ to $2\pi$ decay rates. We get 
from Eq.~(\ref{eq:rates}):
\begin{eqnarray}
&&\left(\frac{\Gamma(3\pi)}{\Gamma(2\pi)}\right)_{X_l}=
\frac{(\cos\theta+\sin\theta)^2}{(\cos\theta-\sin\theta)^2}\cdot
\frac{\langle p_{\omega}\rangle}{\langle p_{\rho}\rangle}\nonumber \\
&&\left(\frac{\Gamma(3\pi)}{\Gamma(2\pi)}\right)_{X_h}=
\frac{(\cos\theta-\sin\theta)^2}{(\cos\theta+\sin\theta)^2}\cdot
\frac{\langle p_{\omega}\rangle}{\langle p_{\rho}\rangle}
\label{eq:ratios}
\end{eqnarray}
 
 Belle attributes all events with $\pi^+\pi^-\pi^0$ mass above $750$~MeV to  
$\omega$ decay and divides by the total number of observed $2\pi$ events. 
They find:
\begin{equation}
\left(\frac{\Gamma(3\pi)}{\Gamma(2\pi)}\right)_{\rm Belle}=0.8
\pm 0.3_{\rm stat}\pm 0.1_{\rm syst}
\label{eq:ratexp}
\end{equation}
 
 The central value is compatible with Eq.~(\ref{eq:ratios}) for:
\begin{equation}
\theta \simeq \pm 20^{0}
\label{eq:theta}
\end{equation} 
for $X_l$ or $X_h$, respectively. Assuming that there are no other significant 
decay modes, the corresponding widths and 
branching fractions for the particle seen in $B^+$ decay are:
\begin{eqnarray}
&&\Gamma=1.6\; {\rm MeV}\;(3.7\;{\rm MeV})\nonumber \\
&&{\cal B}(2\pi)=0.61\;(0.95)
\label{eq:propert}
\end{eqnarray}
where we have listed in parenthesis the properties of the particle 
not seen in $B^+$ decay. The mass difference of the two states is:
\begin{equation}
M(X_h)-M(X_l)=(8\pm 3)\; {\rm MeV}
\label{eq:massdiff2}
\end{equation} 

 We give also the corresponding predictions for the charged state $X^+$, 
which decays via $\rho$-exchange only:
\begin{eqnarray}
&&\Gamma(X^+\rightarrow J/\psi+\pi^+\pi^0)=
\frac{2|A|^2}{8\pi M_X^2}
\langle p\rangle_{\rho}= 4.6 \; {\rm MeV};\nonumber \\
&&{\cal B}(X^+\rightarrow J/\psi+\pi^+\pi^0)\simeq 1 
\label{eq:ratesplus}
\end{eqnarray}

 The value of the mixing angle in (\ref{eq:theta}) is perhaps on the high 
side but still compatible with the general picture. More precise 
data are clearly needed.

 We close this Section by considering the leptonic decays:
\begin{equation}
X(3872)\rightarrow J/\Psi+e^+e^-
\end{equation}

 The lepton pair originates from the coherent superposition of $\rho$ and 
$\omega$ produced in the decay of the $X$. Thus, the branching ratio 
can distinguish between $X_l$ and $X_h$, supplementing the 
measurement of the mass. For simplicity we give the result for the case of
vanishing mixing. A simple calculation gives:
\begin{eqnarray}
&&\frac{d\Gamma(X\rightarrow \psi+e^+e^-)}{ds}= \nonumber \\
&&=\frac{|A|^2 {\cal B}(\rho \rightarrow e^+e^-)}{8\pi M_X^2}
\frac{M_\rho\Gamma_\rho}{\pi}\cdot p(s)\cdot \nonumber \\
&&\cdot \left|\frac{1}{(s-M_{\rho}^2)+i(M_{\rho}\Gamma_{\rho})}\pm
\frac{1/3}{(s-M_{\omega}^2)+i(M_{\omega}\Gamma_{\omega})}\right|^2 \nonumber
\end{eqnarray} 
We have assumed the quark-model ratio for the leptonic amplitudes of 
$\rho$ and $\omega$ and used the narrow width approximation. The sign $\pm$ 
applies to $X_u$ and $X_d$, respectively. Combining with Eq.~(\ref{eq:rates}), 
with $\theta=0$, we find:
\begin{eqnarray}
&&{\cal B}(X_u\rightarrow J/\Psi + e^+ e^-)=0.8\cdot 10^{-4}\nonumber \\
&&{\cal B}(X_d\rightarrow J/\Psi + e^+ e^-)=0.3\cdot 10^{-4}
\label{eq:ratpm}
\end{eqnarray}

\section{Production of $[cq][{\bar c}{\bar q^{\prime}}]$ states in $B$ non-leptonic 
 decays}

 $B^+$ and $B^0$ decays produce superpositions of the two neutral 
states, Eq.~(\ref{eq:Xud}) as well as the charged states:
\begin{equation}
X^+= [cu][\bar d \bar c]; \; X^- = [cd][\bar u \bar c]
\label{eq:Xpm}
\end{equation}
For simplicity, we shall restrict to vanishing mixing.

 We consider first the $B^+$ decay amplitudes for the 
allowed decay:
\begin{equation}
B^+ = ({\bar b}u) \rightarrow {\bar c}+c+{\bar s}+u+({\bar u}+u\; or\; 
{\bar d}+d)
\end{equation}
One additional pair is included in the final state, created from 
the vacuum by the strong interaction.

 If we want a $K$ in the final state, the $\bar s$ must combine either with 
the spectator quark, $u$, to give a $K^+$ (amplitude $A_1$) or with one quark 
from the additional pair, to give either $K^+$ or $K^0$, (amplitude $A_2$). 
Thus we have two independent amplitudes:

\noindent {\bf $B^+$}:
\begin{eqnarray}
&&A(K^+ X_u)=A_1+A_2; \nonumber \\ 
&&A(K^+ X_d)=A_1; \nonumber \\
&&A(K_S X^+)=\frac{A_2}{\sqrt{2}}
\label{eq:Bplus}
\end{eqnarray}

 For $B^0$ decays we have simply to exchange $u$ with $d$ and $K^+$ 
with $K^0$, to get:

\noindent{\bf $B^0$}:
\begin{eqnarray}
&&A(K_S X_d)=\frac{A_1+A_2}{\sqrt{2}}; \nonumber \\ 
&&A(K_S X_u)=\frac{A_1}{\sqrt{2}}; \nonumber \\
&&A(K^+ X^-)=A_2
\label{eq:Bzero}
\end{eqnarray}

\noindent We note, in passing, that these relations follow also from the 
$\Delta I=0$ rule obeyed by the weak transition.

 We noted already that the mass difference 
given in Eq.~(\ref{eq:massdelta}) is larger than the apparent width of the 
$X(3872)$ peak seen by Belle~\cite{x3872}. 
Thus, only one of the two neutral states is produced 
appreciably in $B^+$ decay. Orientatively, we shall assume that:
\begin{equation}
\Gamma(K^+ X_{(u\;{\rm or}\;d)})>4\Gamma(K^+ X_{(d\;{\rm or}\;u)})
\label{eq:ineq1}
\end{equation}
Eq.~(\ref{eq:ineq1}) implies some bound to the production of the 
charged states, 
$X^{\pm}$, not observed thus far. With three amplitudes and two 
parameters, Eq.~(\ref{eq:Bplus}) gives rise to the triangle inequality: 
\begin{eqnarray}
&&|A(K^+ X_u)|+ |A(K^+ X_d)|> \sqrt{2}|A(K_S X^+)| \nonumber \\
&&> ||A(K^+ X_u)|- |A(K^+ X_d)||
\label{eq:ineq2}
\end{eqnarray}
We are interested in the lower bound to the rate of $X^+$, which, due to 
Eq.~(\ref{eq:ineq1}) or Eq.~(\ref{eq:ineq2}), is:
\begin{equation}
|A(K_S X^+)| > \frac{1}{2\sqrt{2}}|A(K^+ X_{q})|
\label{eq:ineq3}
\end{equation}
with $q= u \;{\rm or}\; d$, according to which is the dominant decay product. 

 Eq.~(\ref{eq:ineq1}) has two solutions:

\noindent{\bf $B^+\rightarrow K^+X_{u}$ dominant}:
\begin{equation}
A_1\simeq A_2
\label{eq:udomin}
\end{equation}
or:

\noindent{\bf $B^+\rightarrow K^+ X_{d}$ dominant}:
\begin{equation}
A_1\simeq -\frac{1}{2}A_2
\label{eq:ddomin}
\end{equation}

 We consider now $B^0$ decays, Eq.~(\ref{eq:Bzero}). It is immediate to see 
that 
if $X_u$ dominates $B^+$ decays,  $X_d$ dominates $B^0$ decays and viceversa. 
Thus we are led to predict that
{\it the $X$ particle in $B^+$ and $B^0$ decays are different}, with a 
mass difference given by Eq.~(\ref{eq:massdelta}) 
or (\ref{eq:massdiff2}). In addition, from the corresponding triangle 
inequality, we find:
\begin{equation}
\Gamma(K^+ X^-)> \frac{1}{2}\Gamma(K_S X_q) 
\label{eq:ineq0}
\end{equation}
with $q=u \;{\rm or}\; d$, whichever particle dominates $B^0$ decays.  
Relations~(\ref{eq:ineq3},\ref{eq:ineq0}) 
remain unchanged if one considers an equal mixture 
of $B^0$ and ${\bar B}^0$ decays and adds $K^+$ and $K^-$ events.

 To conclude, we give explicitly the lower bounds to the production of 
$X^{\pm}$ in $B^+$ and $B^0$ decays:
\begin{eqnarray}
&&R^+= \nonumber \\
&&=\frac{{\cal B}(B^+\rightarrow K_S X^+)
\cdot {\cal B}(X^+\rightarrow J/\Psi+\pi^+\pi^0)}
{{\cal B}(B^+\rightarrow K^+ X_{l/h})
\cdot {\cal B}(X_{l/h}\rightarrow J/\Psi+\pi^+\pi^-)}>0.2
\nonumber \\
&&R^0=\nonumber  \\
&&=\frac{{\cal B}(B^0\rightarrow K^+X^-)
\cdot {\cal B}(X^-\rightarrow J/\Psi+\pi^-\pi^0)}
{{\cal B}(B^0\rightarrow K_S X_{h/l})
\cdot {\cal B}(X_{h/l}\rightarrow J/\Psi+\pi^+\pi^-)}>0.53 \nonumber 
\label{eq:xplus}
\end{eqnarray}
to be compared with the upper limit given by 
BaBar~\cite{noXpmbabar}: 
\begin{equation}
R^+ <0.8 \nonumber
\end{equation}
with large errors.

\section{Conclusions}

 The diquark-antidiquark structure explains well the properties of the 
$X(3872)$: spin-parity, narrow width, simultaneous decay into channels 
with different isospin. Taking the $X(3872)$ as input, we have derived 
a spectrum 
which is able to explain spin-parity and decay properties of few other 
particles that could be at odds with a $q\bar q$ picture: $X(3940)$, 
the previously discovered $D_{s}(2317)$ and 
$D_{s}(2457)$ and the $X(2632)$ claimed by SELEX.
 
 Isospin breaking in the wave function and in 
strong decays of these states 
is a distinctive consequence of the asymptotic freedom of QCD, much in the 
same way as narrow widths for heavy quarkonia. Also, all these 
states have to be doublets, unlike the case of the $D-D^*$ molecules, with 
typical mass splittings given by twice the down-up quark mass difference.
The two different states of $X(3872)$ should appear in $B^+$ and $B^0$ 
decays, respectively.

 The crucial test of the scheme, of course, will be the observation of the 
charged or doubly charged partners of the $X$ particles and, more generally, 
the observation 
of heavy states with really exotic quantum numbers. We have derived  
rather strict bounds for the production of $X^+(3872)$ in 
$B$ decays, close to the present limits so that a meaningful test may be 
expected in the near future. The existence of exotic states at low-energy is 
also a pressing issue. 

 The indications derived from the properties of $a/f(980)$ and $X(3872)$ 
seem to us very compelling, so as to warrant a thorough experimental and 
theoretical investigation.

{\sl Acknowledgments}

We would like to thank G.C.~Rossi for interesting discussions on baryonium
and  F. Close, N.A.~T\"{o}rnqvist and the other  
participants 
to the EURIDICE meeting in Barcelona for very useful remarks and comments.
We are indebted to Nando Ferroni and, especially, Riccardo Faccini 
for useful information on the experimental data.

\section*{APPENDIX}

 We give here a simple derivation of the matrix elements of the spin-spin 
Hamiltonian over the $qq-\bar q \bar q$ states. 
We consider states which can be written schematically as:

\begin{equation}
|S_{[cu]}, S_{[{\bar c}{\bar u}]}; J\rangle = |\Gamma, \Gamma^{\prime};
J\rangle=(c^a 
\Gamma_{ab}u^b)({\bar c}^c \Gamma^\prime_{cd}{\bar u}^d)
\label{eq:states}
\end{equation}

$S_{[cu]}$ and $S_{[{\bar c}{\bar u}]}$ are the 
total diquark and antidiquark spin and $J$ the total angular 
momentum. Individual spins are represented by 
$2\times 2$ matrices, $\Gamma^{\alpha}$, with:

\begin{equation}
\Gamma^{0}=\frac{\sigma_2}{\sqrt{2}};\Gamma^i=
\frac{1}{\sqrt{2}}\sigma_2 \sigma^i
\label{eq:gammas}
\end{equation}
for spin $0$ and $1$, respectively. The matrices $\Gamma$ are normalized so 
that:
\begin{equation}
{\rm Tr}[(\Gamma^{\alpha})^{\dagger}\Gamma^{\beta}]=\delta^{\alpha \beta}
\label{eq:norm}
\end{equation}
Spin operators are defined according to:

\begin{equation}
{\bf S}_u|\Gamma\rangle=|\Gamma\frac{1}{2}{\bf \sigma}\rangle; 
{\bf S}_c|\Gamma\rangle=|\frac{1}{2}{\bf \sigma}^T\Gamma\rangle 
\label{eq:spinops}
\end{equation}
Since:
\begin{equation}
{\bf {\sigma}}^T\sigma_2=-\sigma_2{\bf {\sigma}}
\end{equation}
we recover the expected formulae for the total spin operator:
\begin{eqnarray}
&&({\bf S}_u+{\bf S}_c)|\Gamma^0\rangle=0\nonumber\\
&&[(S_u)^i+(S_c)^i]|{\Gamma}^j\rangle=i\epsilon^{ijk}|{\Gamma}^k\rangle
\label{eq:tot}
\end{eqnarray}
We find, also:
\begin{eqnarray}
&&\langle0|{\bf S}_u|1\rangle=-\langle 0|{\bf S}_c|1\rangle
=\frac{1}{2};\nonumber \\
&&\langle1|{\bf S}_u|1\rangle=+\langle1|{\bf S}_c|1\rangle =
\frac{1}{2}\langle1|({\bf S}_u+ {\bf S}_c)|1\rangle
\label{eq:matrixel}
\end{eqnarray}

We can now compute the matrix elements of products of spin operators. We 
have two cases.

\noindent{\bf 1. Same diquark}, e.g. $S_u \cdot S_c$. This operator is just 
a combination of Casimir operators:
\begin{equation}
2({\bf S_u} \cdot {\bf S_c})=({\bf S_{cu}})^2-({\bf S_c})^2-({\bf S_u})^2
\end{equation}
and is diagonal in the basis.

\noindent{\bf 2. Different diquarks}, e.g. $S_u \cdot S_{\bar u}$. We consider 
as an example the $J=0$ states, represented by (summation of repeated
indices understood):
\begin{equation}
|0, 0; 0\rangle=\frac{1}{2}(\sigma_2)\otimes(\sigma_2); |1, 1; 0\rangle=
\frac{1}{2\sqrt{3}}(\sigma_2 \sigma^i)\otimes(\sigma_2 \sigma^i)
\label{eq:zerostates}
\end{equation}

 Using the basic definitions, we have:
\begin{eqnarray}
&&2({\bf S}_u\cdot{\bf S}_{\bf u}) |0, 0; 0\rangle
=\frac{1}{4}(\sigma_2\sigma^i)
\otimes(\sigma_2\sigma^i)=\frac{\sqrt{3}}{2}|1, 1; 0\rangle;\nonumber\\ 
&&2({\bf S}_u\cdot{\bf S}_{\bf u}) |1, 1;0\rangle=\frac{1}{4\sqrt{3}}
(\sigma_2\sigma^i\sigma^j)\otimes
(\sigma_2\sigma^i\sigma^j)=\nonumber \\ 
&&=\frac{\sqrt{3}}{2}|0, 0; 0\rangle- |1, 1;0\rangle
\end{eqnarray}
In conclusion, on the states $|0, 0; 0\rangle$ and $|1, 1; 0\rangle$ we 
obtain the matrices:
\begin{eqnarray}
&&2({\bf S}_u \cdot {\bf S}_c)=\left( \begin{array}{cc}
   -3/2   &  0 \nonumber \\
  0    &   1/2 
\end{array}\right)\nonumber
\end{eqnarray}
\begin{eqnarray}
&&2({\bf S}_u \cdot {\bf S}_{\bar u})=\left( \begin{array}{cc}
   0   &  \sqrt{3}/2  \\
  \sqrt{3}/2  &   -1 
\end{array}\right)
\label{eq:matrices}
\end{eqnarray}

 Using the relations, Eqs.~({\ref{eq:matrixel}), we derive from 
(\ref{eq:matrices}) the representatives of the other spin-spin operators, 
to obtain, e.g. the mass matrix given in Sect. III. 

 We conclude by giving the tensor basis for the $J=1$ states.
\begin{eqnarray}
&&|A\rangle=|0, 1; 1\rangle=\frac{1}{2}(\sigma_2)\otimes
(\sigma_2 \sigma^i); 
\nonumber \\
&&|B\rangle=|1, 0; 1\rangle=\frac{1}{2}(\sigma_2 \sigma^i)
\otimes(\sigma_2); 
\nonumber \\
&&|C\rangle=|1, 1; 1\rangle=\frac{1}{2\sqrt{2}}\epsilon^{ijk}
(\sigma_2 \sigma^j)\otimes(\sigma_2 \sigma^k).
\end{eqnarray}

\end{document}